\documentclass[%
twocolumn,
groupedaddress,
 amsmath,amssymb,
]{revtex4-2}

\usepackage{graphicx}
\usepackage{dcolumn}
\usepackage{bm}
\usepackage{hyperref}

\hypersetup{
    colorlinks=true,
    linkcolor=blue,
    filecolor=magenta,      
    urlcolor=cyan,
    pdftitle={Sharelatex Example},
    bookmarks=true,
    citecolor=blue
}
\begin{document}

\preprint{APS/123-QED}

\title{Reconfigurable Flat Optics with Programmable Reflection Amplitude Using Lithography-Free Phase-Change Materials Ultra Thin Films}

\author{S\'{e}bastien Cueff$^1$}%
\author{Arnaud Taute$^{1,2}$}%
\author{Antoine Bourgade$^3$}%
\author{Julien Lumeau$^3$}%
\author{St\'{e}phane Monfray$^2$}%
\author{Qinghua Song$^4$}%
\author{Patrice Genevet$^4$}
\author{Xavier Letartre$^1$}
\author{Lotfi Berguiga$^5$}
\email{lotfi.berguiga@insa-lyon.fr}
\affiliation{$^1$
 Universit\'{e} de Lyon, Institut des Nanotechnologies de Lyon - INL, CNRS UMR 5270, Ecole Centrale de Lyon - CNRS, 69134 Ecully, France.
}%
\affiliation{$^2$STMicroelectronics, 850 Rue Jean Monnet, Crolles, 38920, France
}%
\affiliation{$^3$Aix Marseille Univ, CNRS, Centrale Marseille, Institut Fresnel, F-13013 Marseille, France
}%
\affiliation{$^4$Universit\'{e} C\^{o}te d’Azur, CNRS, CRHEA, rue B. Gregory, 06560, Valbonne, France}%
\affiliation{$^5$Universit\'{e} de Lyon, Institut des Nanotechnologies de Lyon - INL, CNRS UMR 5270, INSA Lyon - CNRS, 69621 Villeurbanne, France}

\date{\today}

\begin{abstract}
We experimentally demonstrate a very large dynamic optical reflection modulation from a simple unpatterned layered stack of phase-change materials ultrathin films. Specifically, we theoretically and experimentally demonstrate that properly designed deeply subwavelength GeSbTe (GST) films on a metallic mirror produce a dynamic modulation of light in the near-infrared from very strong reflection ($R>80\%$) to perfect absorption ($A>99,97\%$) by simply switching the crystalline state of the phase-change material. While the amplitude of modulation can lead to an optical contrast up to 10$^6$, we can also actively "write" intermediate levels of reflection in between extreme values, corresponding to partial crystallization of the GST layer. We further explore several layered system designs and provide guidelines to tailor the wavelength efficiency range, the angle of operation and the degree of crystallization leading to perfect absorption. 
\end{abstract}


\maketitle


\section{Introduction}
Adjusting the absorption, reflection and transmission properties of systems is the basis of most photonic devices engineering, from mirrors to dispersion gratings as well as photodetectors and solar cells. The recent progress in metamaterials and metasurfaces has provided new methods to precisely control these features to an unprecedented degree. For example, through careful spatial arrangement of dielectric meta-atoms with multipolar resonances we can design metasurfaces tailored for specific optical functionalities. This concept has been exploited to demonstrate flat optics such as lenses, polarizers, retroreflectors, holograms, perfect absorbers, etc. \cite{yu2014flat,wang2018broadband,arbabi2017planar,zheng2015metasurface} that hold promise to surpass the performances of conventional diffractive optics components.

However, both the nanoscale of meta-atoms and the standard materials used for metasurfaces can represent a roadblock for modulation and reconfiguration purposes. Indeed, nanopatterning thin-film materials through etching processes definitively set their geometries and limit their functionalities to a designated purpose. To take a simple example, a TiO$_2$-based metasurface hologram, once fabricated will only display one holographic image, what severely restricts the potential of this technology. 
In that context, recent works demonstrated the potential of phase-change materials (PCM) for tunable nanophotonics. Indeed, this class of materials enables a very large optical modulation at the nanoscale via a fast change of phase in their crystalline structure. 
This large optical modulation of PCMs has been used in specifically designed nanostructures to enable active beam-steerers ~\cite{de2018nonvolatile}, dynamic modulation of light emission ~\cite{cueff2015dynamic}, light absorption ~\cite{carrillo2018reconfigurable} or light transmission ~\cite{howes2020optical}. 

Most of these studies leverage optical frequencies meta-atoms which require in-plane nanostructuration of materials that make their large-scale fabrication and industrial development difficult.
Alternatively, other less technologically constraining methods exist to engineer the optical properties of devices using lithography-free planar thin-films. The most well-known example is the anti-reflection coating, for which the thickness of a transparent thin-film is set at $\lambda$/4$n$ to minimize the reflection through destructive interferences while simultaneously maximizing transmission.
By introducing PCMs in similar engineered thin-films, recent works demonstrated tunable structural coloration \cite{Hosseini2014} or tunable near-perfect absorption \cite{Kats2012,Hendrickson2015a,Hu2017,Wang2018a,Mkhitaryan2017,Sreekanth2018a,Guo2019}. 
So far, these experimental works demonstrated actively switchable optics that are either binary and/or volatile.
However, the complex features of chalcogenide PCMs, and most notably the possibility to actively set them into a state of controlled partial crystallization may be used to a much deeper extent. Indeed their complex refractive index can be encoded into arbitrary intermediate values between those of amorphous and those of fully crystalline. This multilevel crystallization produces stable, non-volatile states that can be driven optically or electrically via pulsed inputs ~\cite{wang20141, zhang2019miniature}. 
Such reconfigurable multilevel optical properties may be exploited as additional degrees of freedom for the design space of multifunctional flat optics with a large number of intermediate states and could be the basis for numerous exciting opportunities in applications such as actively controlled reflectivity modulation, continuous optical power limiting, tunable displays, active spectral filtering and dynamic wavefront shaping.

\begin{figure*}[t]
\centering
\includegraphics[width=12cm]{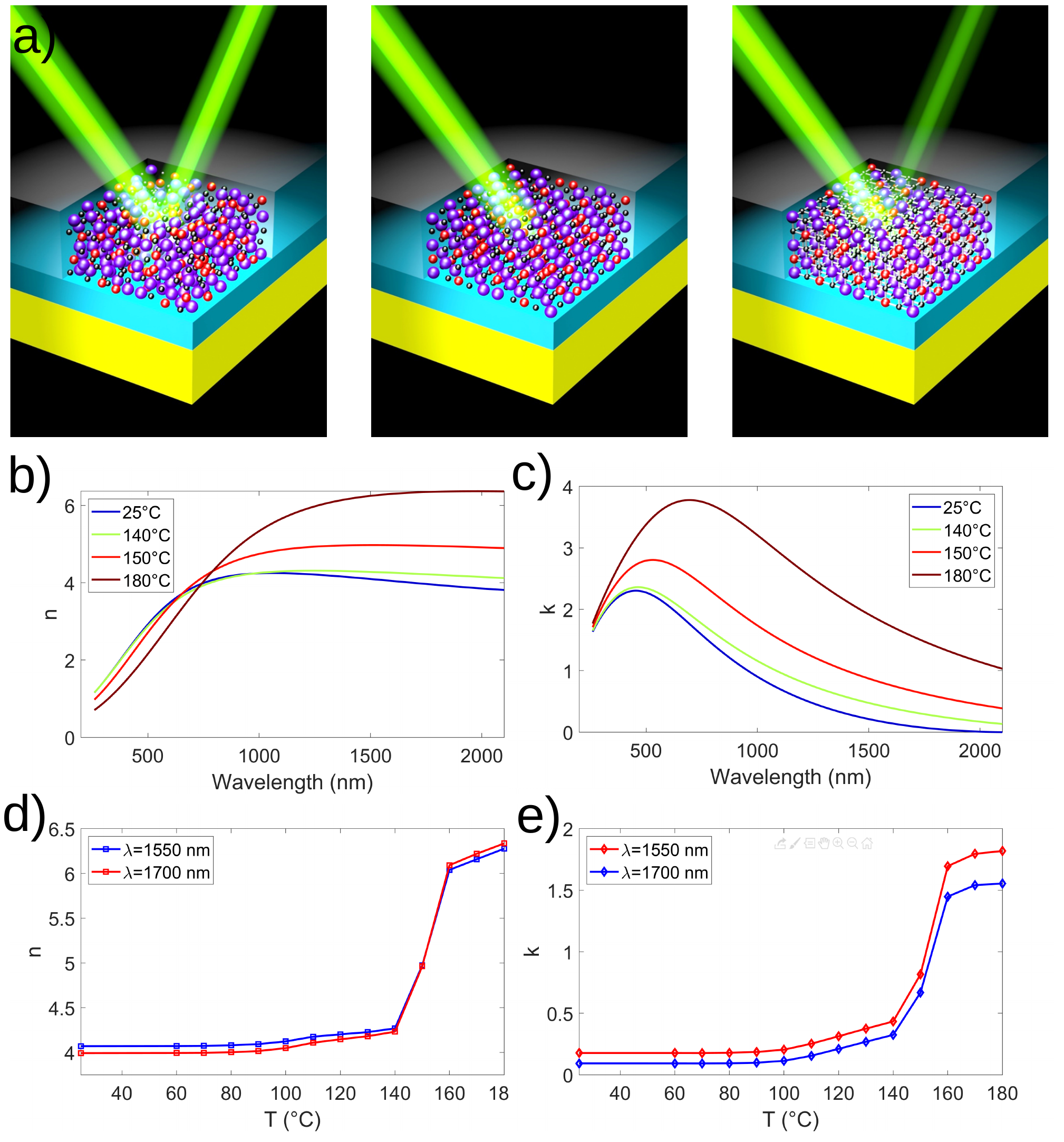}
\caption{a) Principle of the reflectivity modulation when the phase of GST is varied from amorphous to crystalline. The layered system comprises a top GST layer, a thin dielectric layer as a spacer (in blue) and a gold layer (in yellow). From b) to e) experimental dispersion curves of GST as a function of temperature:  b) refractive index $n$ and c) absorption coefficient $k$. in d) Evolution of refractive index and e) absorption coefficient vs temperature at 1550 nm and 1700 nm wavelength.}
\label{Figure1}
\end{figure*}

In this work, we design and demonstrate actively reconfigurable lithography-free flat optics whose optical properties can be continuously tuned from a strong reflection (up to R$\textgreater$80\%) to a perfect absorption (A=(1-R)$\sim$99.99\%) i.e an extinction of -68 dB that is actively controlled by simply adjusting the crystalline fraction of a standard GST thin-film (see Figure \ref{Figure1}a) for an illustration of the concept). Such a modulation depth surpass most of free-space optical modulators reported so far \cite{howes2020optical,gerber1993gaas,sensale2012extraordinary,sun2016surface} with the additional advantage of not requiring any complex nanopatterning processes. By precisely exploiting the non-volatile multilevel states enabled by partial crystallization of GST, we provide comprehensive design rules to simultaneously tailor the wavelength efficiency range, the angle of operation and the degree of crystallization leading to perfect absorption. We further propose practical implementations of this concept in multilayered configurations designed for electrical modulation that conserves all achieved properties.

\section{Optical properties of GST}
Phase change materials such as GST present unique chemical bonding properties (sometimes referred to as resonant bonding or "metavalent bonding") with strong electronic polarizabilities that produce a very large refractive index in the visible and infrared regions, typically in the range of $n\sim$6-7 ~\cite{shportko2008resonant,huang2010bonding,zhu2018unique, raty2019quantum}. This metavalent bonding requires a long range order between atoms and is therefore lost when the material is in an amorphous state. In that latter state, the refractive index falls back to values typical of a semiconductor i.e. $n\sim$4. Changing the crystalline phase of GST from amorphous to crystalline therefore produces an exceptionally large modulation of the refractive index. Furthermore, such a structural reorganization can be driven thermally, electrically or optically. Figure \ref{Figure1}b) and c) illustrate the drastic change of refractive index $n$ and absorption coefficient $k$  of a GST layer of 53 nm deposited on a gold layer upon crystallization. The dispersion curves  $n(\lambda)+ik(\lambda)$ have been measured by ellipsometry (more information in the experimental section) as a function of temperature. Between 120 and 160 $^{\circ}C$, the refractive index $n$ values almost double in the 1200-2000 nm wavelength range upon crystallization, and a transition temperature is found around 130$^{\circ}C$, as shown in the figure \ref{Figure1}d). Simultaneously, the absorption coefficient $k$ also undergoes a drastic change (figure \ref{Figure1}e)). The co-evolution of the real part $n$ and the imaginary part $k$ of the complex refractive index versus temperature follows a quasi-linear variation at all wavelengths. To simplify the numerical calculations, the evolution of $n$ and $k$ as a function of the wavelength in the phase transition range can reasonably be approximated using a linear regression. Using this linear variation $n(k)$ we introduce the quantity $\tau$, that represents the crystallization fraction where 0 and 100 \% correspond to the amorphous and crystalline state, respectively.

\section{Theoretical considerations}

\subsection {Modulation principle}

When a dielectric layer is deposited on a reflective surface, e.g. a thick gold  layer, most of the incident light is reflected back. However, depending on the thickness of this dielectric layer, the reflected light fields from the first interface and second interfaces may either be in phase or in opposite phase. This layer can therefore be used as a means to adjust the reflectivity of the system via constructive or destructive interferences. 
The reflection properties of a generic system comprising a semi-infinite reflective substrate and two thin layers on top, can be expressed as follows:

\begin{equation}
    r=\frac{r_{01}+r_{123} e^{2i \beta_{1}}}{1+r_{01} r_{123} e^{2i \beta_{1}}}
    \label{eq1}
\end{equation}

with $r_{01}$  the Fresnel coefficient at the first interface and $r_{123}$ the effective reflection coefficient of the combined second and third interface. $\beta_{1}$= $\frac{2 \pi}{\lambda}$$\tilde{n_{1}}$$d_{1}$, with $\tilde{n_{1}}$ and $d_{1}$ the complex refractive index and thickness of layer 1, respectively.

From this equation, we can see that there are different ways of tailoring the reflection of such a multilayer stack. The simplest one is to adjust the thickness of the thin-film until obtaining a desired reflection value at a given wavelength. A similar effect can be obtained by choosing a material with appropriate complex refractive index so as to adjust the optical path ($n.d$). Interestingly, such system can be engineered to reach a perfect optical absorption via a mechanism called critical coupling. In equation (\ref{eq1}) this regime can be reached when the numerator equals zero and gives us the following conditions:

\begin{equation}
    R_{01}=R_{123}e^{-4 \pi k_{1} d_{1}/\lambda}
    \label{eq2}
\end{equation}

\begin{equation}
    \Phi_{123}+2\pi n_1\frac  {d_1}{\lambda}-\Phi_{01}=2 \pi m
    \label{eq3}
\end{equation}

where $k_1$ is the wavevector in layer 1, $\lambda$ the wavelength and $\Phi_{01}$ and $\Phi_{123}$ are the phases of $r_{01}$ and $r_{123}$, respectively. Simply put, equations (\ref{eq2}) and (\ref{eq3}) set conditions in terms of amplitude and phase to reach the perfect absorption. Re-arranging equation (\ref{eq2}), we find a more intuitive relation:

\begin{equation}
    R_{01}=R_{123}e^{-\alpha_{1}d_{1}}
    \label{eq4}
\end{equation}

where $\alpha_{1}$ is the absorption in layer 1. Equation (\ref{eq4}) implies that, if the phase conditions of equation (\ref{eq3}) are fulfilled, only two parameters suffice to attain the perfect absorption: the thickness and the absorption of the top dielectric layer. More intuitively, this critical coupling condition corresponds to a regime where the overall radiative losses and absorption losses are equals. This principle has been used to obtained perfect absorption using ultra-thin layers of lossy materials ~\cite{Kats2016}. By introducing a transparent layer between the reflective surface and the top lossy layer we add a supplementary degree of freedom to adjust the reflected phase and reach destructive interferences. This approach has been used by Long et al. \cite{Long2014}. In their work, by tuning $R_{123}$ and $\Phi_{123}$, they provide guidelines to reach the critical coupling conditions in a tri-layer system using various combinations of materials. 
Interestingly, in PCMs one can actively modify the complex refractive index and may therefore dynamically modulate the optical reflection of a stack of thin-films. 
Using PCM as a top layer we not only considerably expand the design space to reach the perfect absorption regime, but we can also optimize such a system to maximize the amplitude of reflectivity modulation. Due to the number of free parameters, in our work we do not use directly equation (\ref{eq1}) but numerical computation of $r$.
The theoretical optical simulations are done with a home-made software based on the transfer matrix method \cite{Yeh2005} and applied in the following work for normal incidence $\theta=0^{\circ}$.

\begin{figure}[htb]
\centering
\includegraphics[width=9cm]{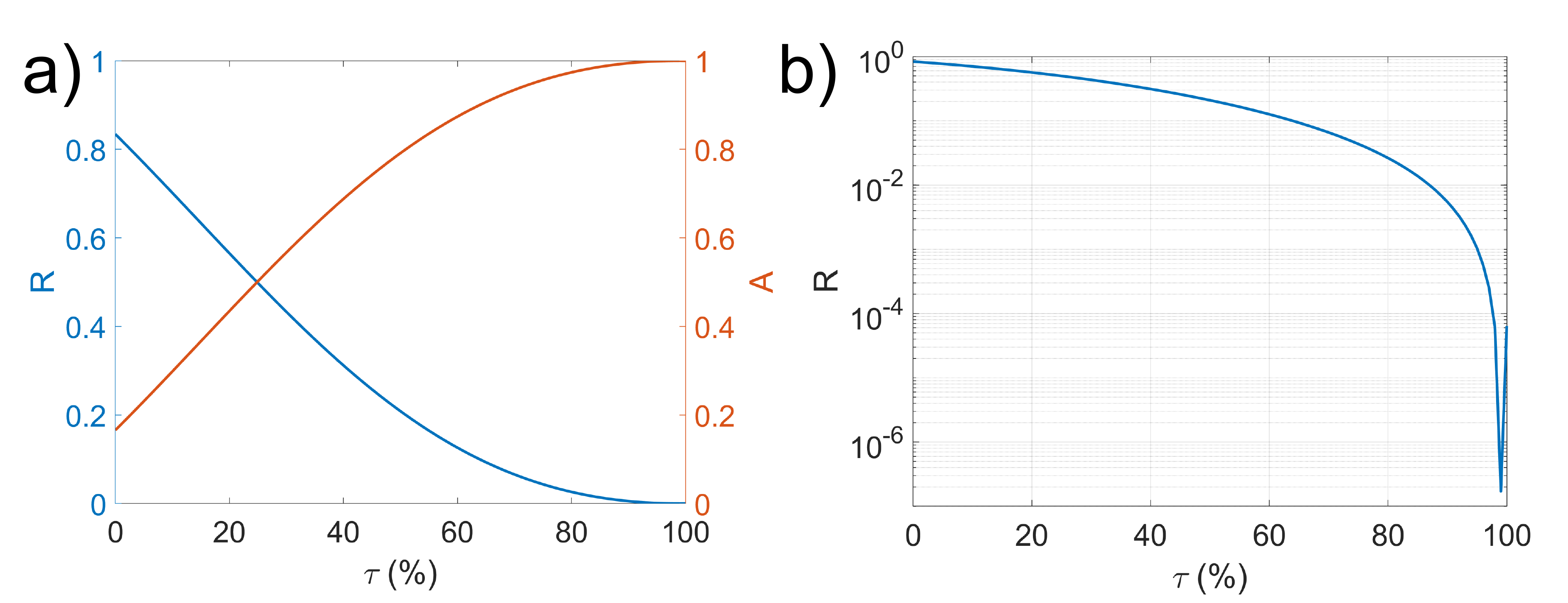}
\caption{Calculated reflectivity at 1550 nm for $Gold/TiO_{2}/GeSbTe$ layer with  thickness 200/67.5/15 nm versus the GST crystallization rate $\tau$ in linear (a) and logarithmic scale (b). The minimum and maximum of reflectivity are  $1.6. 10^{-7}$ and 0.83.}
\label{Figure2}  
\end{figure}

To illustrate this concept, in the following we give an example of a simple planar device that maximizes the reflection modulation. We define a layered system composed, from bottom to top, of Gold/$TiO_{2}$/GST (respectively 200 nm , 67.5 nm and 15 nm). Each thicknesses were calculated to maximize the reflectivity contrast between the amorphous and crystalline states of GST at a given wavelength (here 1550 nm). The calculated amplitude modulation as a function of crystallization is shown in Figure \ref{Figure2}. In the amorphous state (crystallization fraction $\tau=0\%$), we start with a very strong reflectivity R=83 \% and this value is progressively reduced during the crystallization. At the fully crystallized state, we obtain a perfect absorption regime ($A\sim$99.99\%), since the reflectivity reaches a value of 0.000016 \% (i.e -68 dB). In order to give a figure of merit for the amplitude of the reflection modulation, here we define the contrast ratio $C$ between $R_{max}$ and $R_{min}$, the maximum and minimum of reflectivity, respectively, at a given wavelength during the phase change of GST: $C=\frac{R_{max}}{R_{min}}$. In this case C is equal to $5.18\cdot10^{5}$. There could be different ways of quantifying the amplitude of modulation, but this simple figure of merit provides a proper and straightforward comparison between our different designs. This system is very simple but yet very powerful given the large amplitude of modulation we can get by simply changing the phase of GST. Furthermore, properly controlling the crystallization fraction should enable an active access to a desired level of reflection in between the extrema. In other words, we can dynamically prepare this sample to have any arbitrary level of reflectivity between 83\% and 0\% and this value will be kept over time owing to the non-volatile properties of phase-change materials.

In this example and throughout this work, our method to optimize the system is as follows: for Gold/Spacer/GST we compute the reflectvity at a given angle of incidence and wavelength while varying the thicknesses of the two layers above gold. For each couple of thicknesses of Spacer and GST, we compute the reflectivity modulation by varying the crystallization rate from 0 to 100 \% and we extract the two extrema of reflectivty, the contrast ratio C and also the crystallization fraction at which occurs the perfect absorption. By this method the optimal GST/spacer layer thicknesses can be selected providing best absorption and maximal reflectivity i.e highest modulation depth.\\

In the following, we show how we can fully exploit the design space offered by this platform to tailor many features of the modulation, such as the wavelength of operation and the crystallization fraction at which occur the perfect absorption.\\

\subsection{Tailoring the system for a desired wavelength of operation}

As explained in the previous section, for a given 3-layer system (Au/dielectric spacer/GST), the amplitude of reflectivity modulation at a given wavelength can be maximized by choosing an optimized couple of thicknesses for the Spacer/GST system. As shown in Figure \ref{Figure3} b), the progressive change of phase of GST (the crystallization fraction is symbolized by different colors in the graph) produces a $\sim$ -50 dB modulation at $\lambda$=1500nm. By simply adjusting the GST thickness, we can tailor the wavelength at which this optimized contrast occur. This is illustrated in Figure \ref{Figure3} c), where a large contrast $C$ is obtained at different wavelengths for three Au/ITO/GST layered systems. In the following, we define an "8-bit bandwidth" corresponding to the range of wavelength where the contrast $C$ is higher than 256 (i.e. the minimum amplitude modulation needed to encode data via an 8 bit grayscale level). In Figure \ref{Figure3}c),  we show that the respective 8-bit bandwidths for three GST thicknesses 17, 20 and 23 nm, span a wavelength range larger than 150 nm.

Note that the range of validity of this study is limited below 1300 nm due to the increasing absorption of GST but remains valid beyond 1700 nm, throughout the mid-infrared domain. This implies that this system can be designed to operate at arbitrary wavelengths in the IR range, with a contrast ratio higher than 256. Furthermore, by choosing appropriate Au/dielectric spacer/GST thicknesses the bandwidth could be as wide as 300 nm (e.g. with 200/59/302 nm, results not shown here).

Importantly, one may wonder how it is possible to shift the wavelength of operation by 150 nm by simply adding 3 nm of GST to the system. The reason behind this effect is that the crystallization fractions $\tau$ at which occur the minimum and maximum reflectivity are different for each of the three cases considered in Figure \ref{Figure3} c). In the following section, we describe how the system can be designed to operate at specific crystallization fractions.

\subsection{Reaching the perfect absorption at designed crystallization fractions of GST }

\begin{figure}[htb]
\centering
\includegraphics[width=9cm]{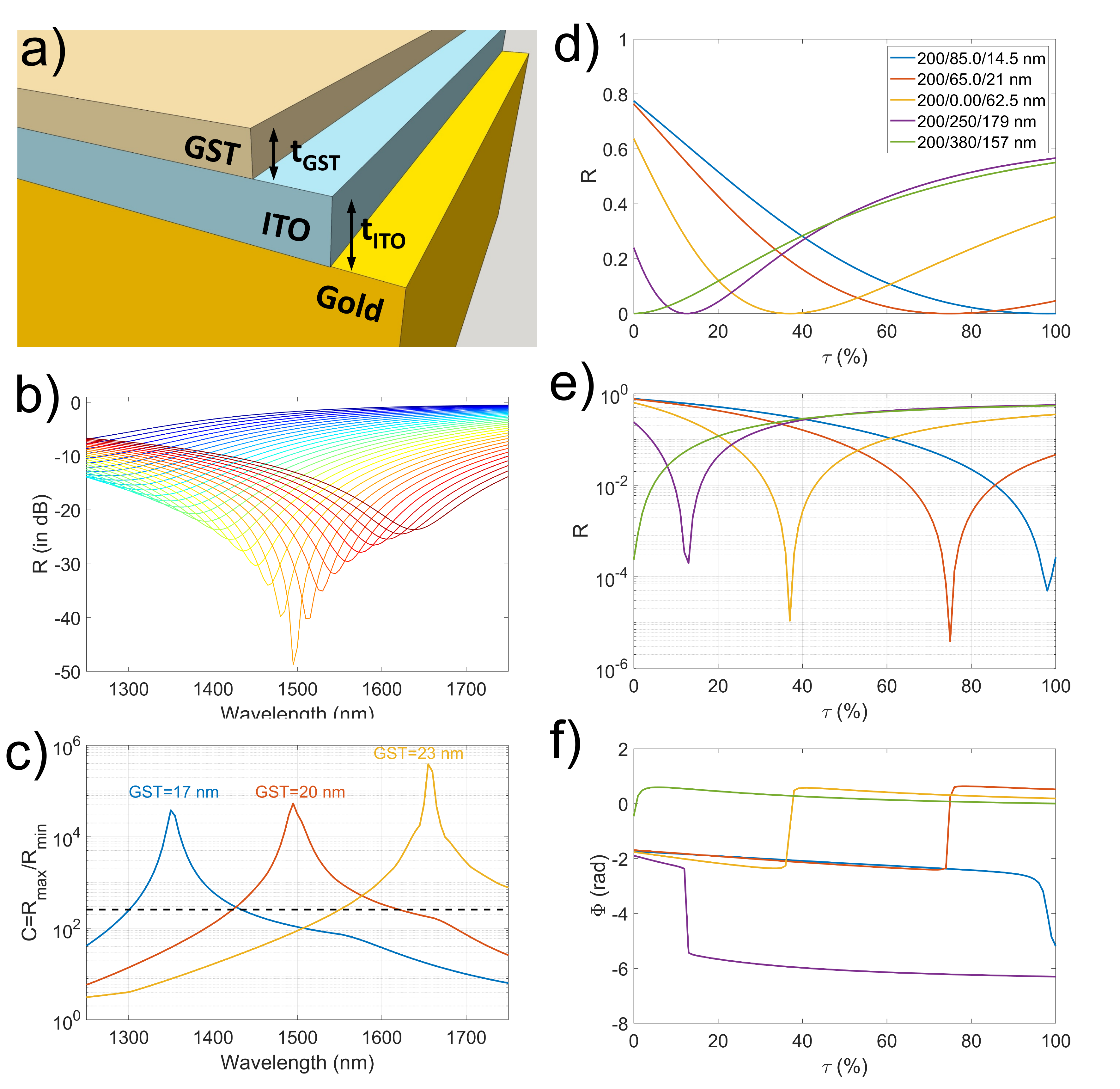}
\caption{a) Sketch of a Gold/ITO/GST system with varying thicknesses. In b) and c) : Tuning the central wavelength by changing the GST thickness in a Gold/ITO/GST layered system. The thickness of gold and ITO are 200 and 65 nm respectively. In b) evolution of the reflectivity spectra vs the crystallization fraction for 20 nm of GST. In c) Evolution of the contrast $C=\frac{Rmax}{Rmin}$ where the GST thickness are 17, 23 and 23 nm for the blue, red, yellow curves.The black dashed curve is the limit for a contrast of 256 (8 bit level of gray). In d), e) and f) tailoring the crystallization fraction for the perfect absorption by adjusting the thickness layers for Au/ITO/GST system. The reflectivity at $\lambda$ = 1550 nm, versus the crystallization fraction in linear scale in (d) and in logarithmic scale in (e). The phase of the reflectivity is shown in (f).
}
\label{Figure3}
\end{figure}

We have seen that it is possible to set PCM layers into states of partial crystallization, hence enabling multilevel intermediate values of complex refractive index. However, most of the reported PCM-based photonic devices are designed to operate as binary devices that simply exploit the fully amorphous and crystalline states. We provided an example of such a device in Figure \ref{Figure2} where the device present a maximal reflectivity when GST is amorphous and a perfect absorption when GST is crystalline, but this was just an illustrative example to highlight the full amplitude modulation of reflectivity one can obtain. With this system, we can indeed go well beyond the sole binary modulation, exploit all intermediate states and select at which crystallization fraction will occur the perfect absorption.
For each layered system shown in Figure \ref{Figure3} c), the perfect absorption is reached at a given value of crystallization fraction $\tau_{pa}$. Using our generic method, we can tailor $\tau_{pa}$ for a given wavelength (here 1550 nm) by adjusting the respective thicknesses of the couple spacer layer/GST layer. This principle is illustrated in Figure \ref{Figure3}d), e) and f) in which the crystallization fraction for perfect absorption is designed to occur at $\tau_{pa}$ = 98$\%$, 75$\%$, 37$\%$, 15$\%$ and 0$\%$ for ITO/GST thicknesses of (85 nm/14.5 nm), (65 nm/21 nm), (0 nm/63,5 nm), (250 nm/179 nm) and (380 nm/157 nm), respectively. It is worth noting that we are able to switch the amplitude  from maximal to minimal reflectivity by either starting from the amorphous or crystalline phase (green and blue curves). The layered system without spacer i.e with GST directly on gold layer is also a solution since it  modulates light with high contrast ratio (yellow curve in Figure \ref{Figure3}d), e) and f)). We should notice also that for every case the maximum of reflectivity is always higher than 50 $\%$ and can reach nearly 80 $\%$ for optimized ITO spacer. Moreover the calculated perfect absorption is always better than -37 dB (see Figure \ref{Figure3}d).

Note that the reflectivity curves $R(\tau)$ shown in figure \ref{Figure3} d) and e) present a minimum of reflectivity but also an abrupt change of optical phase at the minimum, as displayed in \ref{Figure3} f) and in agreement with recent works on tunable perfect absorbers \cite{park2017dynamic,Sreekanth2018a}.\\ 

\section{Experimental demonstration with the Au/GST system}

\begin{figure*}[t]
\centering
\includegraphics[width=16 cm]{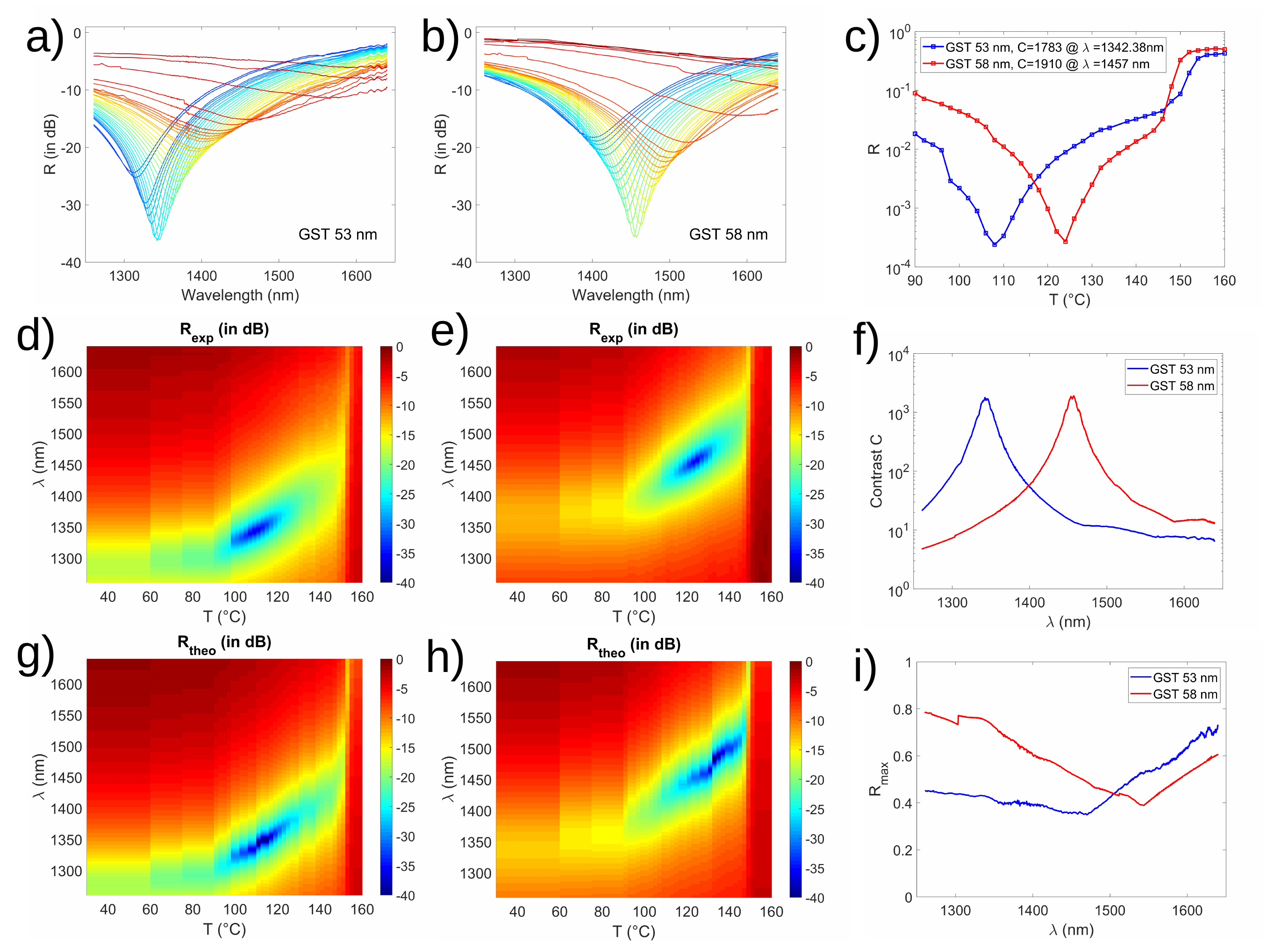}
\caption{Experimental results on the Au/GST layer system and comparison with theory: in a), b), d) and e) experimental reflectivity $10 \cdot Log(R_{exp})$ value vs wavelength and temperature heating for 53 and 58 nm GST layer on 200 nm gold layer. In g) and h) the simulated reflectivity $10\cdot Log(R_{theo})$ vs wavelength and temperature heating for 54.75 and 59 nm theoretical GST layer. In a) and b), the spectra are colored in rainbow jet color code with the increasing temperature from 25$^\circ$C to 160$^\circ$C. In c) the experimental reflectivity curve during phase transition for of 53 and 58 nm GST layers at optimal wavelength 1342 and 1457 nm respectively (i.e. at maximal absorption). In f) The experimental contrast C and Rmax vs the wavelength. In i) The maximal reflectivity for both samples in the wavelength range studied.}
\label{Figure4}
\end{figure*}
To experimentally demonstrate these concepts, we have fabricated Au/GST systems with GST thicknesses of 53nm and 58nm, respectively (more details in the experimental section). 
The modulation of reflectivity is observed for both samples upon thermally-driven crystallization (the reflectivity in dB, i.e $10\cdot Log(R)$, is represented in Figure \ref{Figure4}a) and b)).
With the dispersion curves shown in Figure \ref{Figure1}, theoretical reflectivity spectra $R_{theo}(\lambda)$ for each crystallization rate $\tau$ have been computed and fitted to the experimental  $R_{exp}(\lambda)$ spectra for each temperature $T$. The fitted thickness for GST are found to be 54.75 and 59 nm, in line with the targeted 53 and 58 nm deposited thickness. These fittings (videos shown in supporting information) allow to retrieve the relations between the temperature $T$ and the crystallization rate $\tau$ for each sample (supporting information). Consequently the  theoretical reflectivity $R_{theo}(\lambda,\tau)$ can be converted and represented with temperature scale as $R_{theo}(\lambda,T)$. These retrieved theoretical curves vs the wavelength and temperature are shown in Figure \ref{Figure4}g) and h) and are in full agreement with the experimental ones (in Figure \ref{Figure4}d) and e). A deep absorption is observed around 110$^{\circ}$C and 125$^{\circ}$C before the sharp phase transition for sample 1 and 2 with 53 and 58 nm GST thickness respectively.
As shown in Figure \ref{Figure4}c), the reflectivity modulation for the two GST layer systems 53 and 58 nm is maximal at 1342 and 1457 nm, respectively and reach a minimum of reflectivity of 0.026\% i.e. -36 dB. This translates into an experimental absorption of $A\sim$99.97\%. As expected from our calculations, the thinner GST layer (53 nm) requires a smaller crystallization fraction to reach the perfect absorption and therefore enters that total absorption regime at a lower temperature. At the complete crystallization of GST the reflectivities are maximal and are 43 and 52 \%, respectively. The contrast ratio $C(\lambda)$ versus the wavelength is reconstructed  and shown in Figure \ref{Figure4}f), and the maximal contrast value are  1783 and 1910 for samples with 53 and 58 nm GST thickness (that correspond to 32 and 33 dB, respectively). Moreover, in both cases the contrast is higher than 256 in  the [1318-1367] nm and [1430-1480] nm range i.e. in a bandwidth of 50 nm (see Figure \ref{Figure4}f)).  Inside the bandwidth, the maximal reflectivity is always higher than 43\%.

These results are experimental demonstrations of the different advantages listed previously:  i) a large dynamical modulation from strong reflection to deep absorption with contrasts higher than 10$^3$; ii) a designer wavelength of operation, tailored via the GST thickness; iii) an adjustable crystalline fraction at which occurs the perfect absorption.
 
\section{System with electrodes}

\begin{figure}[htb]
\centering
\includegraphics[width=9 cm]{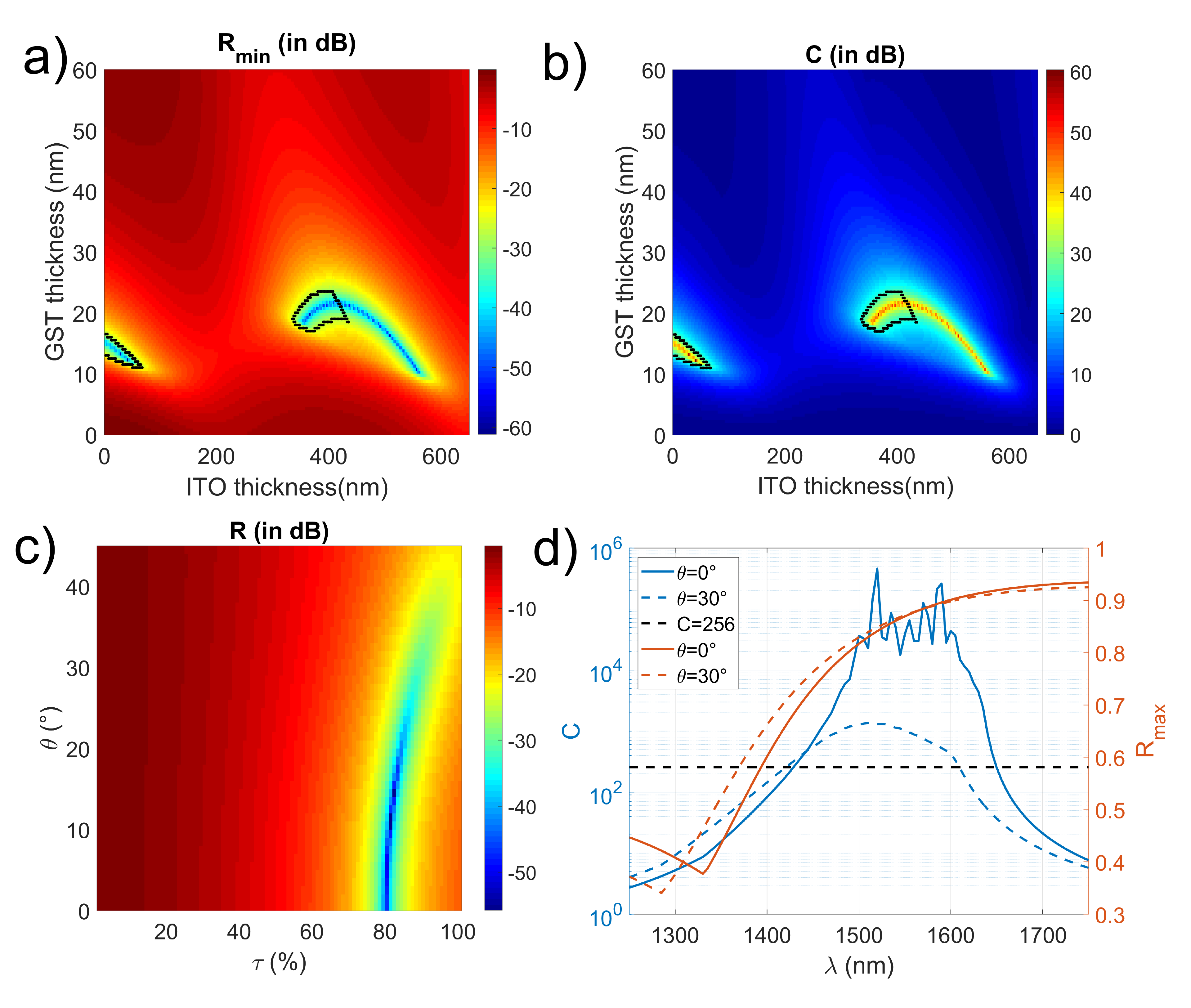}
\caption{Minimum of reflectivity  (in a)) and contrast C (in b)) at 1550 nm upon the change of phase of GST versus the thickness of top ITO and GST in a Au/ITO/GST/ITO layered systems. The Thickness of gold and ITO in contact with gold are 200 nm and 60 nm respectively. The black contour delimits the inside region where the contrast C is higher than 256 and the maximum of Reflection is higher than 50 \%. In c) and d): light modulation and angle of incidence: in c) Simulation of the reflectivity of an Au(200nm)/ITO(60nm)/GST(20nm)/ITO(320nm) layered system versus crystallization rate $\tau$ and angle of incidence $\theta$ at 1550 nm wavelength. In d) Contrast $C$ (blue curves) and maximal reflectivity $R_{max}$ (red curves) at two angle of incidence (0 $^{\circ}$, full line and at 30 $^{\circ}$, dashed line)
}
\label{Figure5}
\end{figure}

We have also computed the reflectivity for a more practical system in which GST is sandwiched between two ITO layers at 1550 nm wavelength. ITO is a transparent conductive oxide allowing the application of an electric field between the two conductive layers or between the top ITO layer and gold layer. This layered system can be a good candidate to modulate light electrically and, in addition, the top ITO layer can serve as a passivation layer to protect GST. For a given thickness of bottom ITO layer (60 nm), we found that there is always a couple of GST, top ITO thickness that provide a large modulation of reflectivity during GST phase variation. It is illustrated in Figure \ref{Figure5}a) at 1550 nm wavelength. The region inside the black contour corresponds to GST and top ITO thickness values where both the maximal reflectivity and contrast are higher than 50 \% and 256 respectively. We can see that absorption with -50 dB can theoretically be reached. For solutions with low top ITO thickness, the maximum of reflectivity is higher than 70 \%. The bottom ITO layer thickness can be arbitrary chosen. However, the reflectivity can be maximized with an optimal bottom ITO thickness which is around 70 nm. Similar results are also observed for the other wavelengths in the 1300-1700 nm range.

\section{Discussion}

Recent works exploited PCMs as a means to modulate the free space optical reflectivity or absorption but with either a limited amplitude of modulation ~\cite{Guo2019,Hu2017,Cao2013} and/or with complicated geometries (Metal-insulator-metal architectures or devices requiring several nanopatterning technological steps ~\cite{howes2020optical,Mkhitaryan2017,wang2014switchable,chen2015tunable,Carrillo2016}). Here, we reveal that there is no need to design and fabricate complex nanostructured devices to reach a very large modulation of reflection/absorption and that a properly designed flat stack of thin-films can produce a dynamical reflectivity modulation with contrast and extinction ratio as high as 10$^6$ and 67 dB, respectively. These figures of merits actually surpass state-of-the-art modulators such as commercially available lithium niobate or liquid crystals free-space modulators, that usually present extinction ratio $\textgreater$10 dB. Our experimental demonstration of a 33 dB extinction ratio also compares favorably to recent experimental works on nanopatterned devices with functional materials such as Graphene or VO$_2$, which reported extinction ratios in the range 10-25 dB \cite{sun2016surface,howes2020optical}.

Furthermore, designer modulation properties, tailored for specific requirements can be obtained by adjusting the thicknesses of both the PCM and the dielectric spacer, providing two degrees of freedom to design the system. The use of a spacer has been reported elsewhere  for binary switch configuration \cite{Hosseini2014,Mkhitaryan2017}, but its role remained elusive and had not been explicitly pointed out. A simplified view would describe the PCM and spacer thicknesses as two separate control sticks to adjust the amplitude and phase of reflected fields, respectively. By adjusting the thickness of the spacer we alter the interference via the introduction of a supplementary phase. On the other hand, by selecting the good thickness of GST layer, it becomes easy to tune the wavelength range  of efficient light modulation. Furthermore, in this work we restricted ourselves to the use of GST, but many other PCMs with various merits are readily available. For example, the emerging large bandgap PCM Sb$_2$S$_3$ could be exploited to expand these designs to the visible range ~\cite{dong2019wide}. The different methods presented here can directly be applied to any other tunable material and should therefore enable demonstrating large reflectivity modulation devices ranging from the visible to the mid-infrared ranges.
Such tailorable properties are much needed, as each given application will require specific crystallization fraction, wavelength of operation and optical contrast. 
Additionally, it is possible to maximize the modulation using both ultra-thin layers of PCMs and low crystallization fraction, as shown in Figure \ref{Figure3}d), e) and f): these two combined features should lead to lower energy consumptions and faster switching times.

Another strong advantage of having flat, deeply-subwavelength layers, compared to patterned or metal-insulator-metal architectures, is the relaxed angle-dependence. So far, all theoretical and experimental results have been presented for normal incidence, but the results remain valid up to an angle of $\pm 30 ^{\circ}$. Modulation and perfect absorption are very good in this range of angles, as shown in figure \ref{Figure5} c) and d). Actually we can see that even at $30 ^{\circ}$ the wavelength bandwidth of large modulation is nearly 200 nm. 
Very importantly, all the methods described in this work can be directly applied to optimize the working conditions around a particular angle (for example $45 ^{\circ}$) simply by adjusting the thicknesses of GST, spacer and potential ITO electrodes.

In view of these different features such as design flexibility, large reflectivity modulation and ease of fabrication, many straightforward applications could be envisioned: flat integrated free space modulator, optical shutter, optical power filter or optical limiter. Using an array of electrodes, this platform could find applications in faster adaptive optics platforms which are used in increasing fields of applications \cite{Popoff2010a,Cizmar2012}. The unpatterned and low thickness nature of the platform makes its fabrication readily accessible to large-scale fabrication for potential integration as dynamically tunable optics in large dimension optical components that are typically used in very large instruments such as telescopes or membrane optics in spacecrafts ~\cite{grezes2018optical,ellerbroek2005adaptive,stamper2000flat}

Finally, we want to emphasize that the unique physical properties of PCMs enable to explore more complex modulation schemes, and in particular by exploiting the multilevel states of partial crystallization. Indeed, in our experimentally shown 33 dB dynamic range modulation of reflectivity, we have been able to set more than 35 distinct states of reflectivity and to reach perfect absorption at an intermediate crystallization state of GST (see Figure \ref{Figure4}). Each of these measured reflectivity level correspond to distinct non-volatile states of partial crystallization. Provided one can precisely and spatially control the local crystalline fraction of PCM, we envision this multilevel non-volatile encoding of reflectivity to be used for a wealth of applications, including grayscale metasurfaces for complex wavefront shaping, reflectivity control devices for attitude control in solar sails \cite{borggrafe2014attitude}  or new kinds of SLMs producing drastically improved holograms' definition compared to the one generated by binary SLM \cite{Goorden2014,Arrizon2005}.

\section {Conclusion}

We have experimentally demonstrated a large optical modulation from a strong reflection to a perfect absorption with a contrast ratio of $\sim2000$ (with measured reflectivity of 0.00026\% i.e -36 dB) using simple unpatterned thin-layers of PCM. We further show that, by adding an appropriate spacer we can tailor : i) the efficient working wavelength conditions throughout the infrared domain, ii) the angle of illumination over all angles, iii) the crystalline fraction of the GST layer at which occurs the tunable perfect absorption. For a fixed layered system thickness, we can obtain a 150 to 300 nm wavelength window enabling modulation with high contrast (higher than 256). Tuning the crystallization fraction where occurs perfect absorption enables a simple binary switch with a very deep absorption modulation (until -68 dB) between the amorphous and crystalline phase. In the other hand, tailoring the crystallization fraction where occurs perfect absorption could be used to modulate light electrically in an ultrathin, low-energy consumption platform via a grayscale level scheme.

\begin{acknowledgments}
This work is partly supported by the French National Research Agency (ANR) under the project SNAPSHOT (ANR-16-CE24-0004).
\end{acknowledgments}

\appendix
\section{Supporting Information}  
Supporting Information is available from the following link:\href{https://mycore.core-cloud.net/index.php/s/IOHqnoklxQlWtyG}{Link2SupportingInformation}.

\section{Experimental Section}
\subsection{Samples Fabrication}
We start by depositing a 200 nm gold layer (e-beam evaporation) on a silicon substrate, followed by depositing a GST layer. GST layers were obtained by using a Bühler SYRUSpro 710 machine associated with an OMS 500 optical monitoring system. Granules made with stoichiometric Ge$_2$Sb$_2$Te$_5$ were placed into a Mo liner. A focused electron beam was then used to heat-up the material with typical current of a few tens of mAmp. Specific e-beam pattern was developed in order to secure uniform evaporation of the material. Samples were placed onto a rotating calotte situated at a distance of about 600 mm from the crucible to allow achieving layers with good uniformity over the substrate aperture. Deposition were carried out at room temperature and GST was evaporated at a rate of 0.25 nm/s that was controlled with a quartz crystal microbalance. A relative precision of the thickness better than 1 nm and an absolute precision within 2\% was achieved using this technique. Finally, previous expertise ~\cite{joerg2016optical} has shown that by adapting properly the deposition parameters, it is possible to keep a composition close to that of a raw material, securing that the properties of the initial material will be maintained. We fabricated two samples with two different GST thicknesses: the first (sample 1) with 53 nm and the second one (sample 2) with 58 nm.\\
\subsection{Measurements of GST Optical Properties}
The GST samples were optically characterized between 260 and 2100 nm using a Horiba Jobin-Yvon spectroscopic ellipsometer. To study the crystallization of the GST layers, samples were placed on a heating stage for 5 minutes before being allowed to cool down to room temperature. The ellipsometry spectrum is then measured at room temperature, to avoid crystallization during the measurement. This process is repeated using the same sample and increasing the temperature of the stage by 10$^{\circ}$C steps, starting from 60$^{\circ}$C up to 180$^{\circ}$C. The different spectra are then individually fitted using a model comprising a GST layer following a Tauc-Lorentz dispersion formula. This enables the extraction of the refractive index and extinction coefficient of the GST for each intermediate crystallization fraction from amorphous to fully crystalline.\\
\subsection{Reflection Measurements}
For reflectivity measurement the light of a broadband superluminescent LED is sent on the GST side with a 5X M plan apo NIR Mitutoyo objective. The focused spot diameter at the sample is around 30 $\mu m$ and the cone angle  of illumination is around 3$^{\circ}$. The reflected light is sent to an optical spectrum analyser (Anritsu MS9740 A) and the spectrum is recorded in the 1250-1650 nm range. The sample is heated from 90$^{\circ}$C to 160$^{\circ}$ by increasing the temperature every 2 minutes by  step of 2$^{\circ}$C  . At the end of every temperature step a  reflectivity single spectrum is recorded. The reflected spectra are all normalized to the reflection spectrum of a silver mirror in order to accurately measure the absolute reflectivity of the sample  $R_{exp}(\lambda,T).$\\

\bibliography{Biblio_PCM}

\end{document}